\def\year{2017}\relax
\documentclass[letterpaper]{article} 
\usepackage{aaai20}  
\usepackage{times}  
\usepackage{helvet} 
\usepackage{courier}  
\usepackage[hyphens]{url}  
\usepackage{graphicx} 
\usepackage{graphicx}  
\usepackage{subfigure}
\usepackage[ruled,lined,linesnumbered,algonl]{algorithm2e}
\usepackage{multirow}
\usepackage{booktabs}

\title{Short Text Classification via Term Graph}
\author{\Large \textbf{Wei Pang}\\ 
School of Artificial Intelligence, \\Beijing University of Posts and Telecommunications\\ 
pangweitf@\{bupt.edu.cn,163.com\} 
}
\begin{document}

\maketitle

\begin{abstract}
Short text classification is a method for classifying short sentence with predefined labels. However, short text is limited in shortness in text length that leads to a challenging problem of sparse features. Most of existing methods treat each short sentences as independently and identically distributed (IID), local context only in the sentence itself is focused and the relational information between sentences are lost. To overcome these limitations, we propose a PathWalk model that combine the strength of graph networks and short sentences to solve the sparseness of short text. Experimental results on four different available datasets show that our PathWalk method achieves the state-of-the-art results, demonstrating the efficiency and robustness of graph networks for short text classification.
\end{abstract}

\section{Introduction}

Short textual sentences are produced in an explosive way in recent years, such as user comments\cite{Expansion17} on shopping website, movie reviews, search query\cite{Expansion17} for web search engine, and rapidly growing bullet screen \cite{BulletScreen}, or named $danmu$\cite{DanMu} message on video website. Understanding these short text has become an important problem for a variety of applications. However, short text is, as the name implies, shortness in text length, a typical sentence-level textual data. Compared with document-level text data, short text lacks of context and sufficient word occurrences, since its shortness, and hence it often leads to their feature space is very sparse. Besides, it can also lead to a challenge problem of robustness, since in real world application, small perturbations of short sentence features are more likely to result in misclassification. In this paper, we aim to alleviate the problem of sparsity and robustness for short text classification by using graph network at the word level.

\begin{figure} \centering
\includegraphics[width=0.40\textwidth]{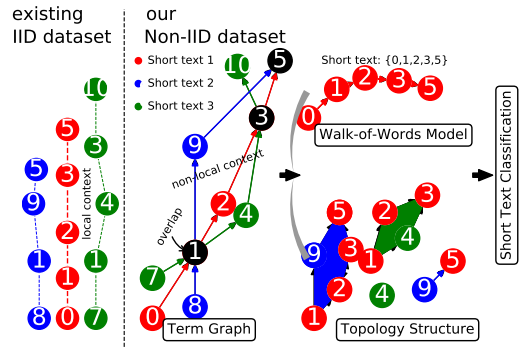}
\caption{An idea of short text classification via term graph. The left is IID data, local context only in the sentence is focused. The right is Non-IID data, context is captured not only in the sentence itself but also in the graph networks. Richer contextual information can be used to short text classification.}\label{fig_overview}
\end{figure}

Traditional classification methods for document-level text heavily rely on rich feature space \cite{JOINT18,Knowledge17}. Commonly used models are bag-of-words (BoW) \cite{Internal09} or N-grams \cite{Manning12} and tf-idf term weighting scheme. On the one hand, owing to short text has very few words, existing methods often fail to provide sufficient features \cite{Internal09,Knowledge17}. On the other hand, conventional machine learning methods always treat dataset as independently and identically distributed (IID), as well as deep learning methods, leading to the BoW model only capture local contextual information, as the left of Fig.~\ref{fig_overview} illustrated, this IID assumption neglect the underlying correlations among sentences. A more serious problem is that existing classifiers are vulnerable to adversarial perturbations \cite{Adversarial18}, for example, given an negative $danmu$ message, if we add one or more trivial words that can't change its sentiment, it is possible to misclassify as positive message. 

Enriching short text is an effective approach by introducing external knowledge\cite{Internal09,Knowledge17,Words15N,Auxiliary16}. Given short sentences, their external information can be mined from the returned snippet of search engine\cite{VERYFEW12}, or knowledge bases\cite{Internal09,Knowledge17}, such as Wikipedia and WordNet\cite{Internal09,VERYFEW12}. The challenge to expend short text is tend to topic drift\cite{LIHANG14,Expansion17}, or the introduced features might be ambiguous\cite{Internal09,Knowledge17,LIHANG14,Expansion17}. 

However, in our view, there exists considerable redundancy information between the sentences. For example in task of sentiment analysis, the sentences that have the same sentiment, may share similar paraphrase or synonymy words in common. In particular, suppose that there are three short sentences are crossed at the word level, as shown in Fig.~\ref{fig_overview},  where the circled number denote a word, they have the same context words while only few keyword or phrase is different, it may be redundant information, we can utilize these information to argument each other, and these couplings between sentences tend to reflect intrinsic characteristics in datasets. Based on these observations, we treat text datasets as non-IID, aiming to make full use of internal relations in training corpus.

Underlying the non-IID assumption, we propose a PathWalk algorithm for short text classification. It consists of two textual representation models and one classification method. 

First, for representing the sentences corpus, the whole corpus is converted into a directed graph, called term graph. Where  the node correspond to a word, directed edge indicate the order of words in a sentence. Moreover, edge with direction can be used to model a special word order in pair of words, since inversion of a pair of words may lead to semantic changes. For example a term graph in Fig.~\ref{fig_termgraph} that show complex relations between sentences, the number of edges is much more than the number of nodes, and intuitively, these edges could provide rich non-local contextual information for short text, which is the basis of our proposed PathWalk algorithm.

Second, for representing a short sentence, we present a walk-of-words model that use a sequence of nodes and directed edges in term graph to denote a sentence.

Third, we propose a method (i.e. PathWalk) that sample some non-local contextual information in term graph and use them to classify short text. In term graph, sentences that are crossed with each other form many closely connections, such as three nodes forms a triangular, four nodes forms a polygon, as shown in Fig~\ref{fig_overview}. To distinguish the types of non-local contextual information in term graph, we adopt the topological feature, see more details in later.

In the end, the key contributions of our work are:
\begin{itemize}
\item We consider the sentence corpus as Non-IID dataset in place of typical IID, aims to capture non-local contextual information between sentences. 
\item We propose a term graph that have the capacity to represent the whole sentence corpus by a directed graph. It establishes the relationship between sentence and sentence and help alleviate the problem of data sparsity.
\item We propose a walk-of-words model that can denote a sentence by a sequence of nodes and directed edges, instead of the typical bag-of-words model.
\item We propose a PathWalk method that enable classify short textual sentence efficiently, and our model achieves new state-of-art results.

\end{itemize}
 
\section{Models}

Here a short text relate to the length is limited to only dozen words\cite{Survey14}, such as comment reviews, search query and DanMu message. We start by describing some concepts. Graph networks of words is defined as following.

Term Graph, is a directed graph $G = \{V, E\}$, consist of nodes and directed edges, where $V$ is a set of nodes, correspond to the vocabulary of training corpus, a node represent a word, $E$ is the set of directed edges, which represent a preorder relation between words, the arrow points from current word to its followed from left to right in a sentence. 

As seen in Fig.~\ref{fig_termgraph}, where the edges have direction that represent a certain relation between pairs of words, the node and edge are both embed into a low-dimensional vector space, denoted $\overrightarrow{v}$ and $\overrightarrow{e}$ respectively. Intuitively underlying term graph, we assume that the edges and nodes together play a role for expressing semantics of sentence, e.g., different word order usually convey different meanings, in this case, the edge embedding encode information on word co-occurrence. To our knowledge, we first use the edge as independent feature to capture the relationships among words.

Local context is only inside the sentence itself because of a sentence is a word sequence in a linear way. However, term graph is not linear, the non-local context can be described as:

Non-local context, based on term graph, the contextual information can be introduced via many directions from network neighborhood, not only include local context inside the sentence itself but also include external context in the graph networks.

In this work, we aim to jointly two different types of applications, network embedding and text classification, learning classification-oriented, low-dimensional vector representation for nodes and edges that can be used for short text classification.

\subsection{Sentence Corpus Representation} 

\begin{figure}[htb] \centering
\includegraphics[width=0.40\textwidth]{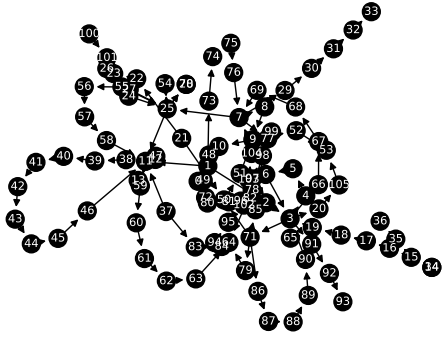}
\caption{A real term graph, is built from the training dataset at word level, where every circled number denote a single word, the directed edge means word appears in sequential order.} \label{fig_termgraph}
\end{figure}

Existing methods usually view the sentence corpus are IID, one of limitations is that the rich relationships among sentences are lost. We transform the textual corpus into a term graph, as illustrated in Fig.\ref{fig_termgraph}. Particularly, the number of nodes is smaller than the number of edges, there exist a lager links in term graph, and these links provide rich internal connections among words. Therefore, different from previous works, we adopt graph networks of words as our basic data structure for short text classification.

\subsection{Short Sentence Representation}

Walk-of-words model:  A short text sentence is represented as a sequence of alternating nodes and directed edges in term graph.

\noindent For example, a short sentence $S$ of length $5$ is composed of: $S = \{0,(0,1),1,(1,2),2,(2,3),3,(3,5),5\}$ in Fig.~\ref{fig_overview}, where nodes means a single word and word appears in succession define a directed edge, the edge with direction can also encode special information about word order, such as the order of edge $(1, 2)$.

\begin{figure}[htb] \centering
  \subfigure[]{
    \label{fig_topology:a}
    \includegraphics[width=0.22\textwidth]{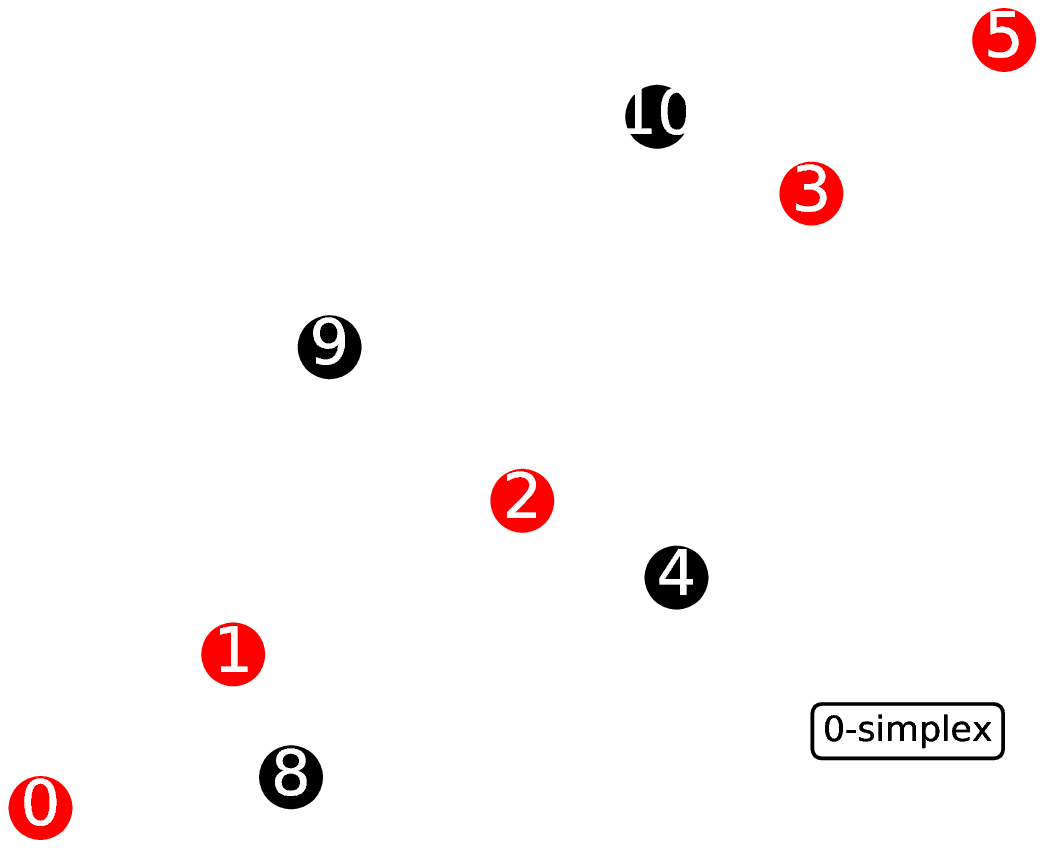}} 
   \hspace{0.00in}
    \subfigure[]{
    \label{fig_topology:b}
    \includegraphics[width=0.22\textwidth]{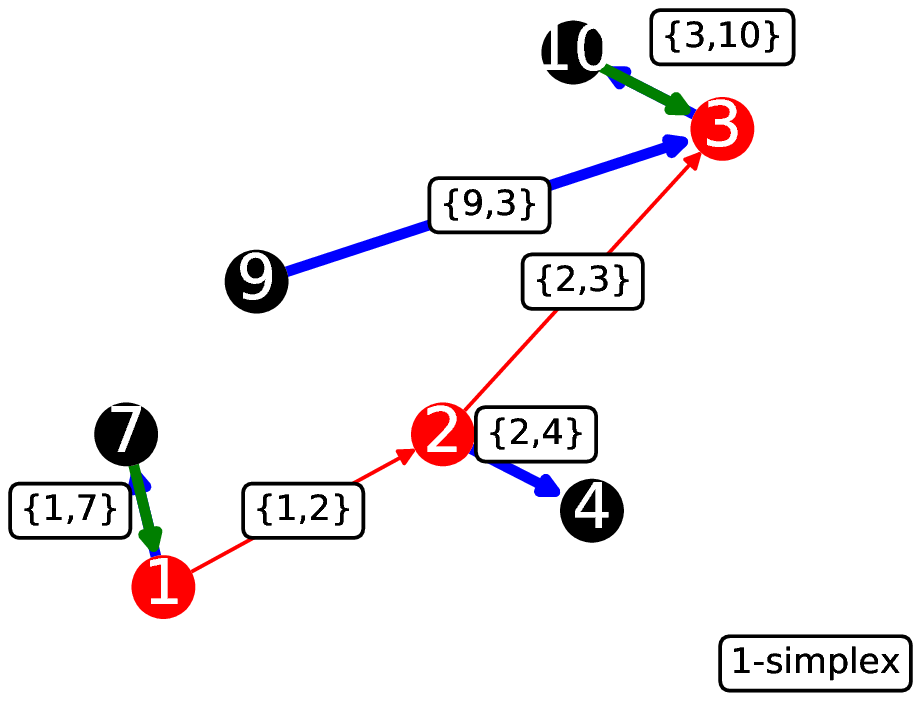}}
   \vspace{0.00in}
    \subfigure[]{
    \label{fig_topology:c}
    \includegraphics[width=0.22\textwidth]{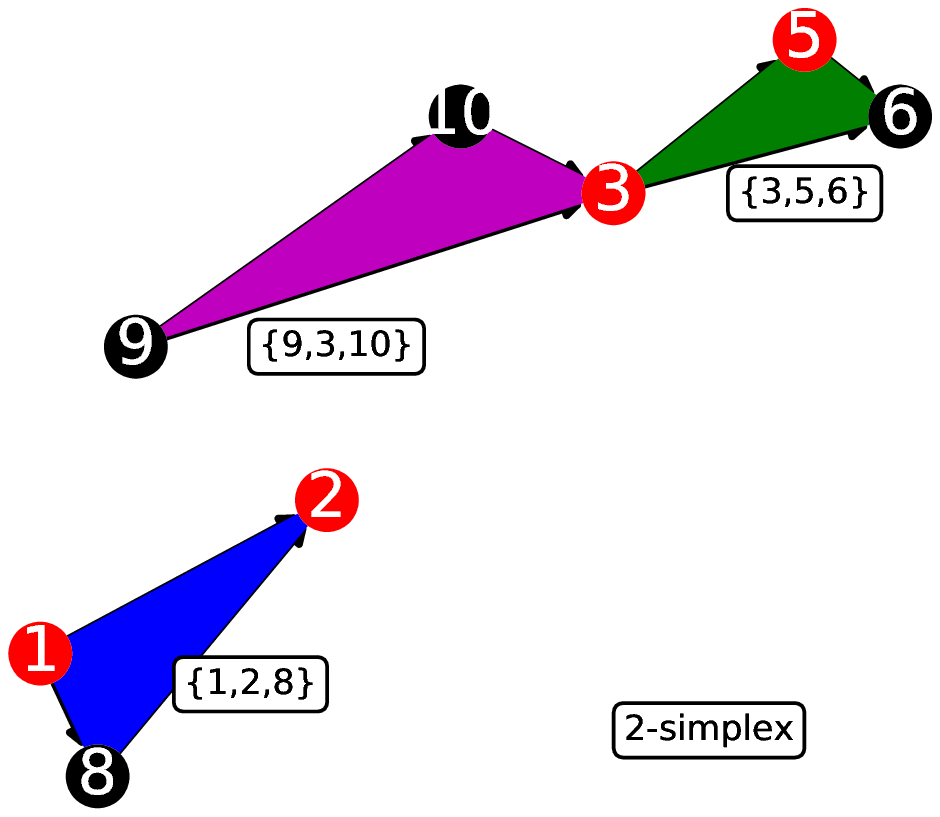}}
   \hspace{0.00in}
    \subfigure[]{
    \label{fig_topology:d}
    \includegraphics[width=0.22\textwidth]{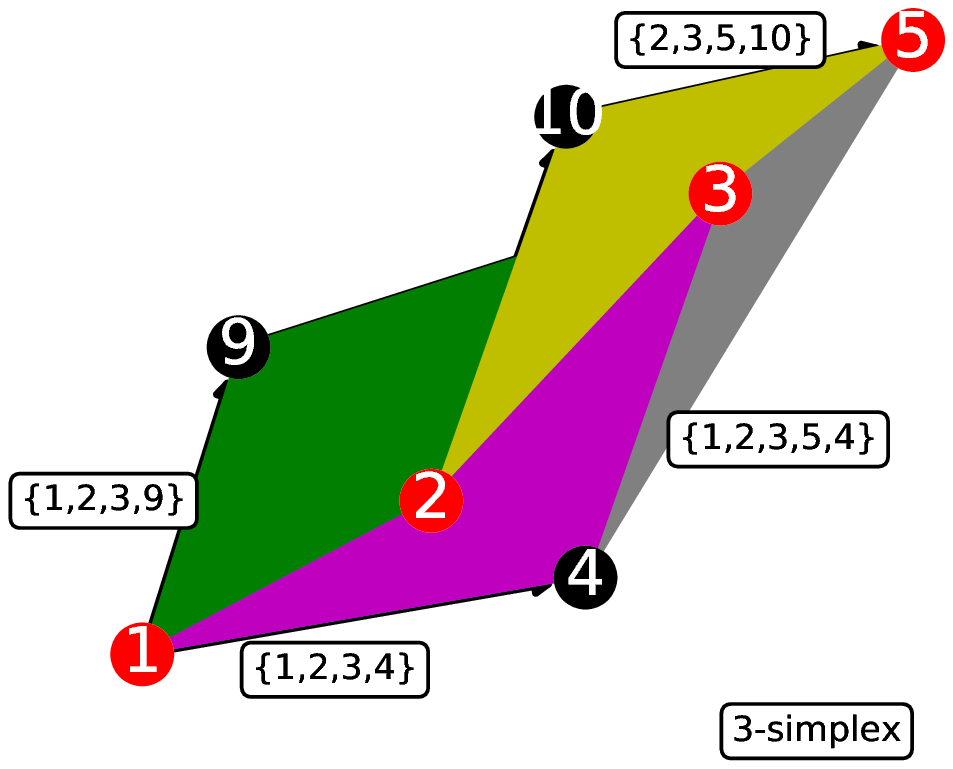}}
    \caption{Four basic types of topological structure as features used for short text classification: a, $0$-simplices, such as single words in sentence. b, $1$-simplices, i.e. the directed inner-links between two single words is encoded as independent features. c, $2$-simplices, such as triangle $\{1,2,8\}$. d, $3(4)$-simplices, such as quadrangle or polygon show in figure.} \label{fig_topology}
\end{figure}

\subsubsection{Network Topology} 
\begin{figure}[htb] \centering
  \subfigure[Bag-of-words model]{
    \label{fig_shortextmodel:a}
    \includegraphics[width=0.22\textwidth]{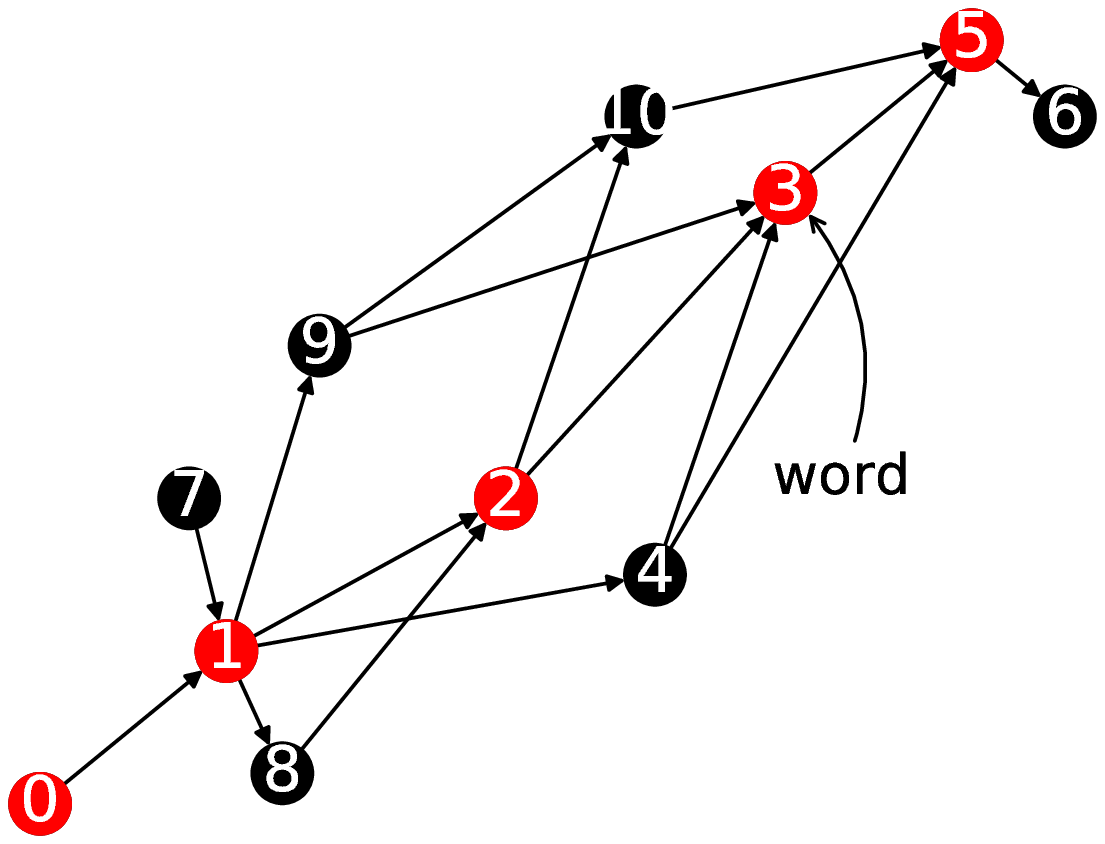}} 
   \hspace{0.00in}
    \subfigure[Walk-of-words model]{
    \label{fig_shortextmodel:b}
    \includegraphics[width=0.22\textwidth]{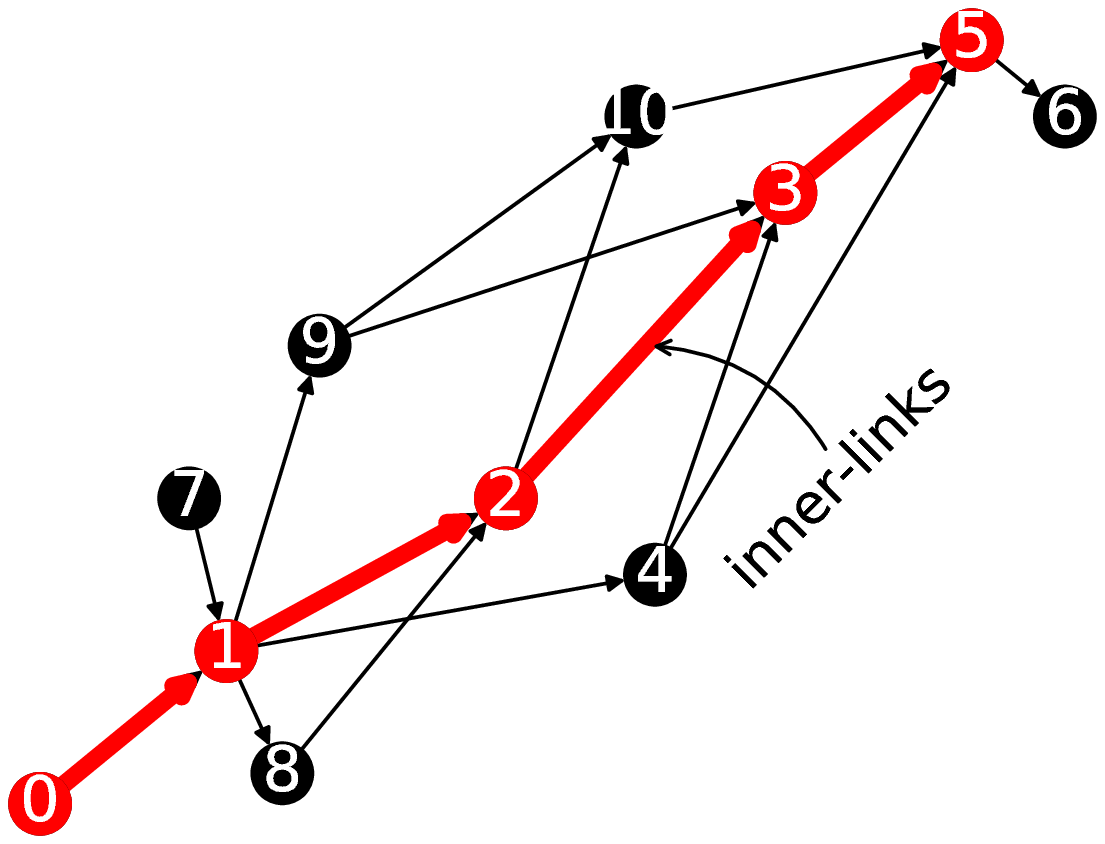}}
    \vspace{0.00in}
    \subfigure[Walk-of-words model plus topological edge]{
    \label{fig_shortextmodel:c}
    \includegraphics[width=0.22\textwidth]{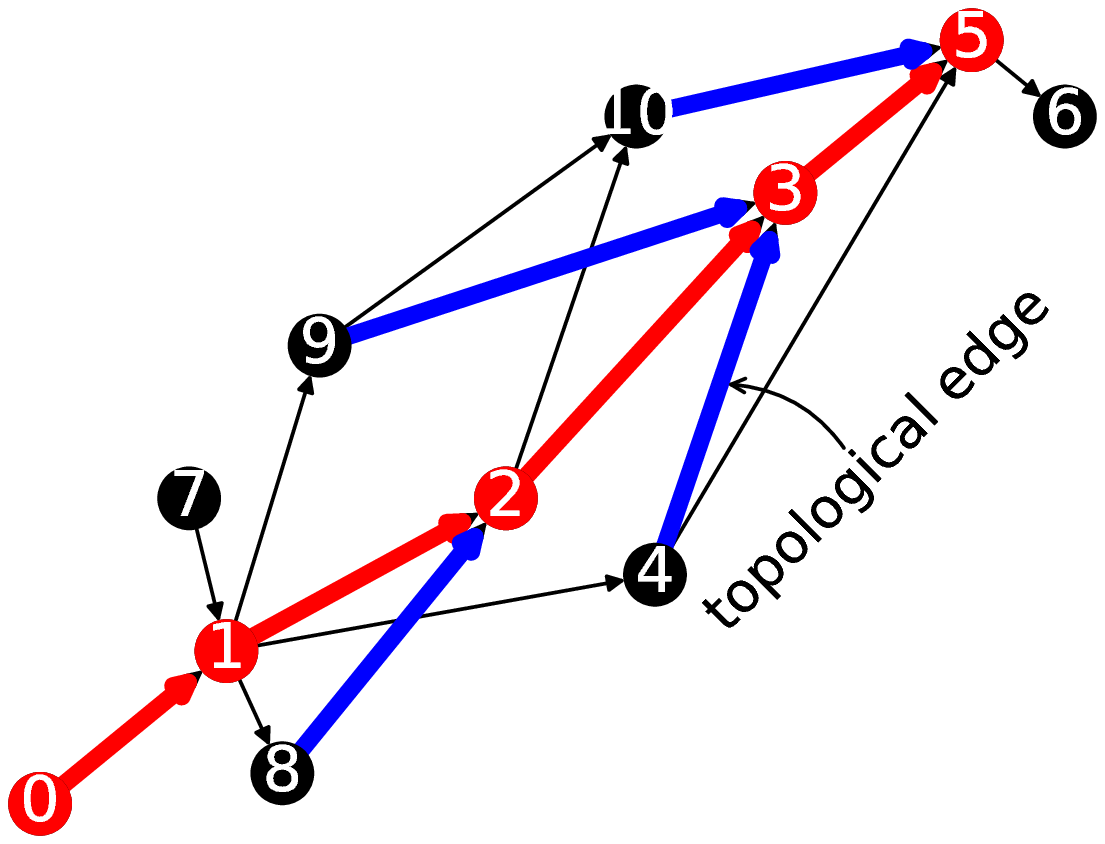}}
    \hspace{0.00in}
     \subfigure[Walk-of-words model plus topological edge and node]{
    \label{fig_shortextmodel:d}
    \includegraphics[width=0.22\textwidth]{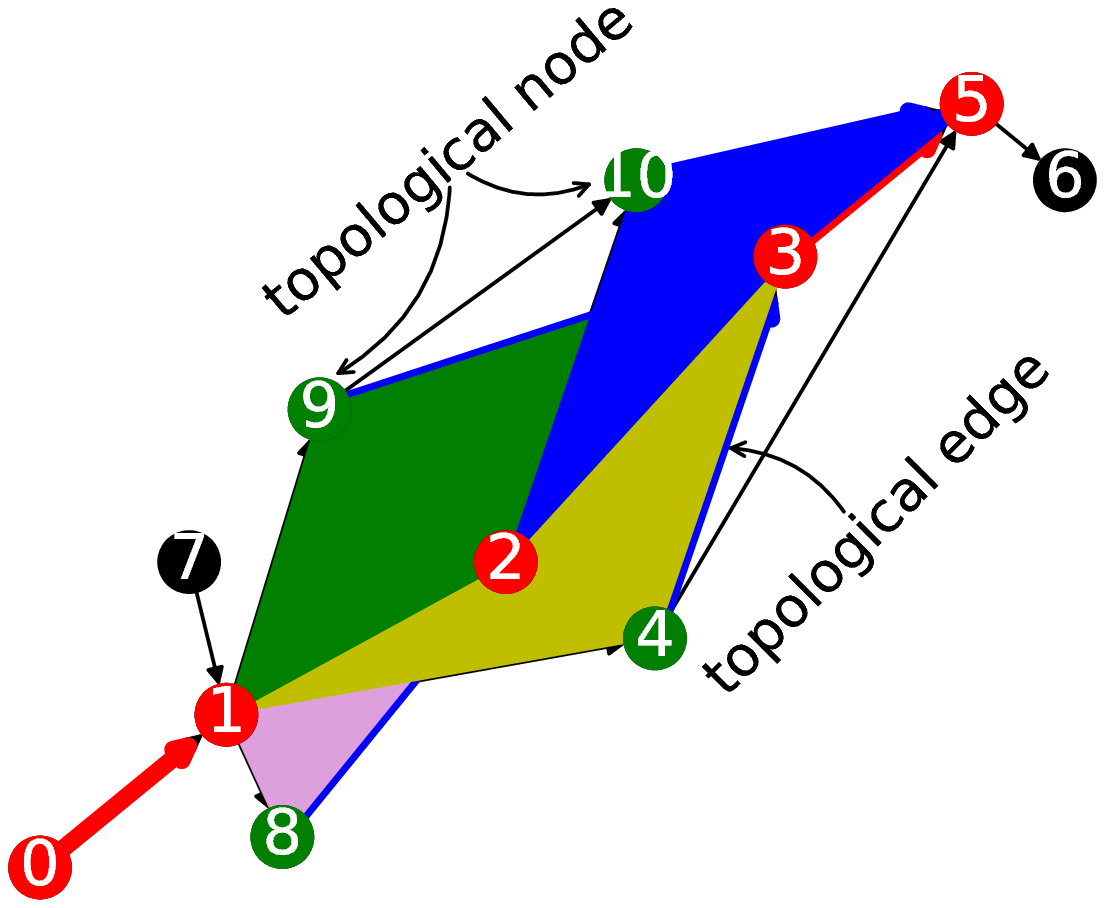}}
    \caption{Four types of short text representation on graph, take sentence $S$ for illustration: a, Bag-of-words model ($BoW$), let $BoW = \{0, 1, 2, 3, 5\}$, then $S = BoW$. b, the proposed Walk-of-Words model, let $E_{in} = \{(0, 1), (1, 2), (2, 3), (3, 5)\}$, then $S = BoW \cup E_{in}$. c, let $E_{t} = \{(4, 3), (8, 2), (9, 3), (10, 5)\}$, then $S = BoW \cup E_{in} \cup E_{t}$. d, let $N_{t} = \{8, 4, 9, 10\}$, then $S = BoW \cup E_{in} \cup E_{t} \cup N{t}$.} \label{fig_shortextmodel}
\end{figure}

For a given short sentence, how to draw valid topological features relative to it. Take a short sentence $S$ as example in Fig.~\ref{fig_shortextmodel:a}, the nodes $\{1, 2, 3, 5\}$ are overlapped with neighbor sentences. The basic idea is that the two overlapping sentences might form a simple local triangle or higher-order polygon, these structure are frequently appears in term graph, called network topology. Moreover, the nodes and edges exist in topology that belong to external sentence might provide useful information for $S$. Even more evocatively, the semantic information can be flowed along with directed edges that make fully utilize data itself. To this end, we define four types of local network topology as contextual features, as see in Fig.~\ref{fig_topology}.

Underlying Non-IID assumption, the topological structure we expect to capture the interaction among training samples that have overlapped nodes. According to Fig.~\ref{fig_topology}, we define four basic types of simplical complex (network topology) as below, which can be used as contextual information for short sentence $S$ to reduce the problem of sparsity. 
\begin{itemize}
\item $0$-simplex, a set of single nodes shown in Fig.\ref{fig_topology:a}, node denote a single word, which corresponding to bag of single words model.
\item $1$-simplex, two node with the inner directed edge between them, such as $\{9, 3\}$ in Fig.\ref{fig_topology:b}, note that the ordered edge can be used to encode the word order. 
\item $2$-simplex, a simple triangle simplicial complex see in Fig.\ref{fig_topology:c}. A $2$-simplex like the blue triangle $\{1,2,8\}$ show in Fig.~\ref{fig_topology:c}.
\item $3(4)$-simplex, visualized in Fig.~\ref{fig_topology:d}, a $3$-simplex shape such as quadrangle $\{1, 2, 3, 4\}$, a $4$-simplex shape like pentagon $\{1, 2, 3, 5, 4\}$.
\end{itemize}

Importantly, in order to avoid introducing unnecessary noise, we give a constraint for network topology. Take a $2$-simplex $\{1, 2, 8\}$ as example in Fig.~\ref{fig_shortextmodel:c}, there is only one node $\{8\}$ comes from neighbor sentence, $\{1, 2\}$ must in $S$, the node $\{1\}$ point to $\{8\}$ but immediately return to self node $\{2\}$ of $S$. We use this simple triangle as a feature w.r.t. node $\{2\}$. Ideally, node $\{8\}$ or edge $(8, 2)$ might bring contextual information, which is introduced to capture short-range context between $\{1\}$ and $\{2\}$.

The $3$-simplex $\{1, 2, 3, 4\}$ seen in Fig.~\ref{fig_shortextmodel:c}, is considered to have the ability to capture long-range dependencies between $\{1\}$ and $\{3\}$ in $S$. Another case is a $4$-simplex $\{1,2,3,5,4\}$ in Fig.~\ref{fig_topology:d}, $\{1, 2, 3, 5\}$ is a snippet of sentence $S$, although the first word $\{1\}$ is far away from the last $\{5\}$, but obviously the first have another path available to the last via external node $\{4\}$ that not in $S$. This shortcut plus node $\{4\}$ are considered as contexts of $S$, enhancement to semantic connection within $S$, in other words, they can offer richer contextual information to each other. This is why the Non-IID assumption works.

\subsubsection{Integrated Walk-of-Words model with Topological Structure} In this subsection, we combine topological feature and walk-of-words model for representing short text. We describe three types of short sentence representation on top of term graph, visualized in Fig.~\ref{fig_shortextmodel}.

First, Bag-of-words Model. A short sentence $S$ is only a bag of words, ignore the order of words, shown in Fig.~\ref{fig_shortextmodel:a}. This representation is widely used in traditional machine learning tasks.

Second, our Walk-of-Words Model. Bag-of-words model plus with inner-links, let $E_{in}$ denote the inner-links, shown in Fig.~\ref{fig_shortextmodel:b}, adding the internal edges of $S$. Different from the previous RNN-base text representation, we use the directed edge as an independent feature. In details, $S = \{0, (0, 1), 1, (1, 2), 2, (2, 3), 3, (3, 5), 5\}$, is a sequence of alternating nodes and edges.

Third, Walk-of-Words Model with topology features. As see in Fig.~\ref{fig_shortextmodel:c}, let $E_{t}$, $N_{t}$ denote topological edge and topological node respectively. We initially use topological edges that marked in the figure, as external features for expanding short text $S$. Furthermore, another higher-order topologies, visualized in Fig.~\ref{fig_shortextmodel:d}, such as a $2$-simplex $\{1, 2, 8\}$, $3$-simplex $\{1, 2, 3, 9\}$, $\{1, 2, 3, 4\}$ and $\{2, 3, 5, 10\}$, a $4$-simplex $\{1, 2, 3, 5, 4\}$. The topological nodes that not in $S$, such as $\{8,9,10,4\}$, can also be used as external contexts to expand short text $S$, in order to capture non-local contextual information and provide much semantic features for short text representation.

\section{PathWalk: Short Text Classification Algorithm}
\begin{figure}[htb] \centering
\includegraphics[width=0.40\textwidth]{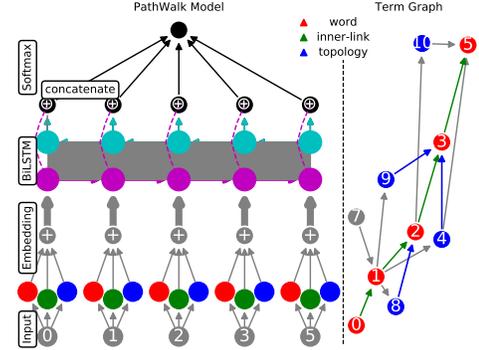}
\caption{The model architecture of the proposed PathWalk.} \label{fig_architecture}
\end{figure}

In this section, we present the PathWalk method for short text classification on top of term graph. The architecture is illustrated in Fig.~\ref{fig_architecture}, consists of four major structures. 

First, Input layer, extracting topological features from term graph, including topological edges and topological nodes, those features are marked by blue color at the right of Fig.~\ref{fig_architecture}. Besides, Walk-of-Words representation correspond to the alternating red nodes and green edges.

Second, Embedding layer, which maps node and edge into dense vectors.

Third, BiLSTM layer, a one-layer bidirectional long short term memory (BiLSTM) encoder.

Last one is an output layer, we use SoftMax as our basic classifier.

According to Fig.~\ref{fig_shortextmodel}, each short text representation corresponds to a feature space, obviously, from Fig.~\ref{fig_shortextmodel:a} to Fig.~\ref{fig_shortextmodel:d}, the number of features used are gradually increasing. However, we would like to verify whether these contextual features can help to improve classification, Fig.~\ref{fig_shortextmodel:a} as our baseline, so that we design three variants of PathWalk algorithm for testing.


\subsection{PathWalk-$\uppercase\expandafter{\romannumeral1}$ method}
According to Fig.~\ref{fig_shortextmodel:b}, based on walk-of-words model, a sentence $S$ of length $n$, its feature space can be defined as:
\begin{equation}
\overrightarrow{S}_{1:n} = \sum_{i=1}^{n} \overrightarrow{v}^{(i)} + \overrightarrow{E_{in}}^{(i)},
\end{equation}
where $\overrightarrow{S}_{1:n}$ denote the sentence representation of $S$, by summing of it's embedding vectors, $\overrightarrow{E_{in}}^{(i)}$ is the $i$-th inner-links embedding shown in Fig.~\ref{fig_shortextmodel:b}, $\overrightarrow{v}^{(i)}$ denote the $i$-th node embedding. To simply the model, we directly use the summing of node vector and edge vector. The objective function is shown in Eq.~\ref{equ_loss}:
\begin{equation} \label{equ_loss}
\min_{\overrightarrow{v},\overrightarrow{e}} \frac{1}{M}\sum_{i=1}^{M} L(f(\overrightarrow{S}^{(i)}), y^{(i)}) + \lambda l2\_loss
\end{equation}
where $y$ denotes the class of sentence $S$, $l2\_loss$ is a regularization that we employ half the $L2$ norm, $\lambda$ is the learning rate, $f$ denote a SoftMax function that is applied to the output of BiLSTM module and then convert it into probabilities.

\subsection{PathWalk-$\uppercase\expandafter{\romannumeral2}$ method}
According to Fig.~\ref{fig_shortextmodel:c}, the topological edges are introduced to expand the feature space of $S$. Note that for each node of $S$, it has several topological edges, they all can be used as contextual information, in practice, we only select the topological edge of size $K$. The sentence representation is defined as:
\begin{equation}
\overrightarrow{S}_{1:n} = \sum_{i=1}^{n}\{\overrightarrow{v}^{(i)} + \overrightarrow{E_{in}}^{(i)} + \sum_{j=1}^{K}\overrightarrow{E_{t}}_{j}^{(i)}\},
\end{equation}
where the symbol $\overrightarrow{E_{t}}$ is the embedding of topological edge.

\subsection{PathWalk-$\uppercase\expandafter{\romannumeral3}$ method}
According to Fig.~\ref{fig_shortextmodel:d}, we calculate the higher-order topology to be used as additional features, such as the triangle or quadrangle covered with color shown in Fig.~\ref{fig_shortextmodel:d}. In this case the sentence representation can be written as:
\begin{equation}
\overrightarrow{S}_{1:n} = \sum_{i=1}^{n}\{\overrightarrow{v}^{(i)} + \overrightarrow{E_{in}}^{(i)} + \sum_{j=1}^{K}\overrightarrow{E_{t}}_{j}^{(i)} + \sum_{j=1}^{K}\overrightarrow{N_{t}}_{j}^{(i)}\},
\end{equation}
where $\overrightarrow{N_{t}}$ is the embedding of topological node, such as the green nodes $\{8,4,9,10\}$ in figure. Similar to topological edge, we also sample $K$ topological nodes.

\subsection{Random Sampling} The node in-degree follows a power-law-like shape distribution, that means the size of topological feature is varying in a range. To be fair, we randomly sample topologies of size $K$ as contextual features for every node of $S$.

\subsection{The PathWalk algorithm}

The pseudocode of three PathWalk variants is shown in Algorithm~\ref{algopathwalk}. First step, scanning each training sentences to construct term graph, where the directed edge from previous word point to current word; preprocessing each sentence using walk-of-words representation and sampling topological features; then optimizing the model using $Adam$ algorithm, these steps are performed in sequential.

\begin{algorithm*}

\SetKwFunction{WalkofWord}{WalkofWord}
\SetKwFunction{ToDirectedGraph}{ToDirectedGraph}
\SetKwFunction{SampleTopologyLinks}{SampleTopologyLinks}
\SetKwFunction{SampleTopologyNodes}{SampleTopologyNodes}
\SetKwFunction{SumEmbedding}{SumEmbedding}
\SetKwFunction{PadSequences}{PadSequences}
\SetKwFunction{Model}{Model}

\caption{PathWalk algorithm} \label{algopathwalk}
\KwData{Training dataset $D$}
\KwIn{Fixed sentence length $l$, Number of topologies $K$, Dimensions $d$}
\KwOut{model, Node and Edge Embedding}
\BlankLine
$G$ = \ToDirectedGraph($D$)\;
modle = \Model($d$)\;
\BlankLine
\For{epoch $\leftarrow 1$ \KwTo $20$}{
	\If{PathWalk-$\uppercase\expandafter{\romannumeral1}$}{
		batch = \WalkofWord($G$, $D$)\;
	}
	\ElseIf{PathWalk-$\uppercase\expandafter{\romannumeral2}$}{
		batch = \WalkofWord($G$, $D$)\;
		$\overrightarrow{oe}$ = \SampleTopologyLinks($G$, batch, $K$)\;
		batch = \SumEmbedding(batch, $\overrightarrow{oe}$)\;
	}
	\ElseIf{PathWalk-$\uppercase\expandafter{\romannumeral3}$}{
		batch = \WalkofWord($G$, $D$)\;
		$\overrightarrow{oe}$ = \SampleTopologyLinks($G$, batch, $K$)\;
		$\overrightarrow{\tau}$ = \SampleTopologyNodes($G$, batch, $K$)\;
		batch = \SumEmbedding(batch, $\overrightarrow{oe}$, $\overrightarrow{\tau}$)\;
         }
	batch = \PadSequences(batch, $l$)\;
	model.AdamOptimizer(batch)\;
 }
\end{algorithm*}

\section{Experiment and Evaluation}
\subsection{Datasets}
To evaluate effectiveness of the proposed PathWalk method, we conduct experiments with four different domain datasets, detailed statistics are summarized in Table~\ref{tab_datasets}.

\begin{itemize}
\item RT-Polarity data(for short, RT)$\footnote{http://www.cs.cornell.edu/people/pabo/movie-review-data/}$, a popular movie reviews dataset v$1.0$ for sentiment analysis with positive/negative labels.
\item SST\cite{SST13}, a Stanford Sentiment Treebank dataset contains fine grained sentiment labels, such as very negative, negative, neutral, positive and very positive. In our work we do not use the neutral class, and merging the very negative and negative data to negative dataset, similar, merging the positive and very positive dataset to positive dataset for binary sentiment classification.
\item Video Query dataset(aka Query), Chinese queries dataset come from a video website, each query has a topic, either music or sport category. This dataset is used to binary topic classification task.
\item Chinese DanMu dataset(aka DanMu), user reviews dataset, we gather from social media sites and give emotion tagging of positive or negative, and then used to binary sentiment classification problem.
\end{itemize}

\begin{table}[htb]
\caption{Dataset statistics. Num., the number of datasets; Nodes, Edges, the number of nodes and edges in term graph at single word level; Lang, the language of text; AvgLen(std), average sentence length with the standard deviation.} \label{tab_datasets}
\begin{tabular}{l|c|c|l|c|c}
\toprule
Dataset&Num.&Lang&AvgLen(std)&Nodes&Edges\\
\midrule
DanMu&50k&CH&6.47(5.49)&3981&76,321\\
RT&10k&EN&22.36(\textbf{13.37})&18,862&106,752\\
SST&10k&EN&19.27(9.23)&18,302&88,109\\
Query&100k&CH&8.40(4.12)&5,290&209,085\\
\bottomrule
\end{tabular}
\end{table}

\subsection{Compared Methods and Setup}
 
In \cite{CNNWord2vec}, the author proposed a shallow CNN model for sentence classification TextCNN, which achieve state-of-the-art results\cite{Expansion17} and contains only one convolutional layer followed one max pooling layer, use three filters of size ${3,4,5}$ for capturing contextual information. fastText is another state-of-the-art\cite{BagofTricks, Expansion17} baseline for text representation and classification, one spot is to use subword embedding to overcome sparsity. A strong baseline method is a typical one-layer BiLSTM follows a SoftMax function\cite{BiLSTM18}. We use fixed sentence length of $10$ for Chinese sentence while of length $20$ for English sentence, padding to fixed length with zero vector or truncating where necessary. We set the mini-batch size of $128$, embedding size of $128$-dimension for both node and edge, as well as the size of BiLSTM hidden states. We implement our experiments using $Tensorflow$ and use $Adam$ to optimize the model. Note that we preserve all the words and punctuation in dataset, these settings are common for all methods.

\subsection{Parameter Sensitivity w.r.t. $K$}

\begin{figure}[htb] \centering
  \subfigure[]{
    \label{fig_K_Sensitivity:a}
    \includegraphics[width=0.22\textwidth]{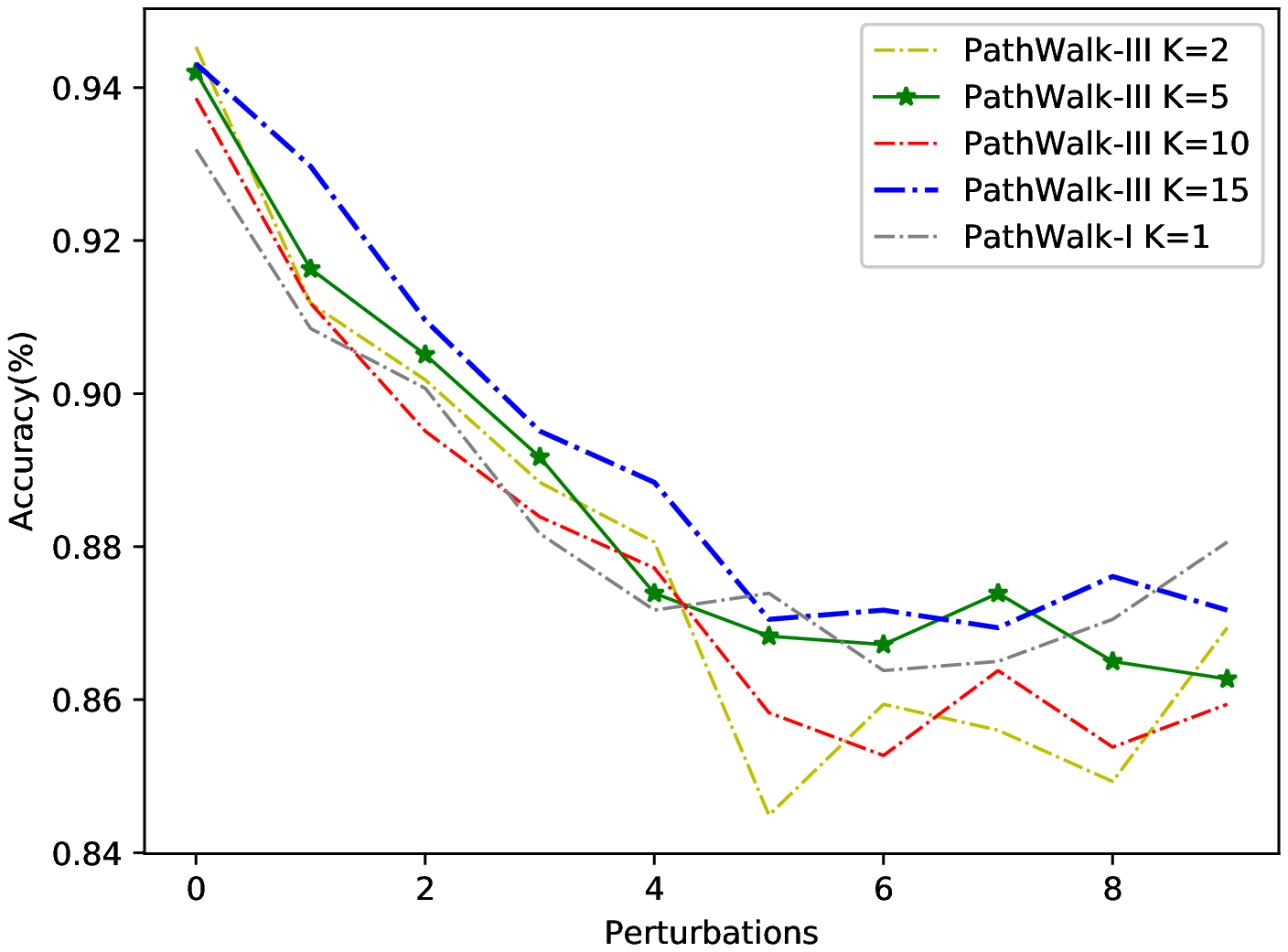}} 
   \hspace{0.00in}
    \subfigure[]{
    \label{fig_K_Sensitivity:b}
    \includegraphics[width=0.22\textwidth]{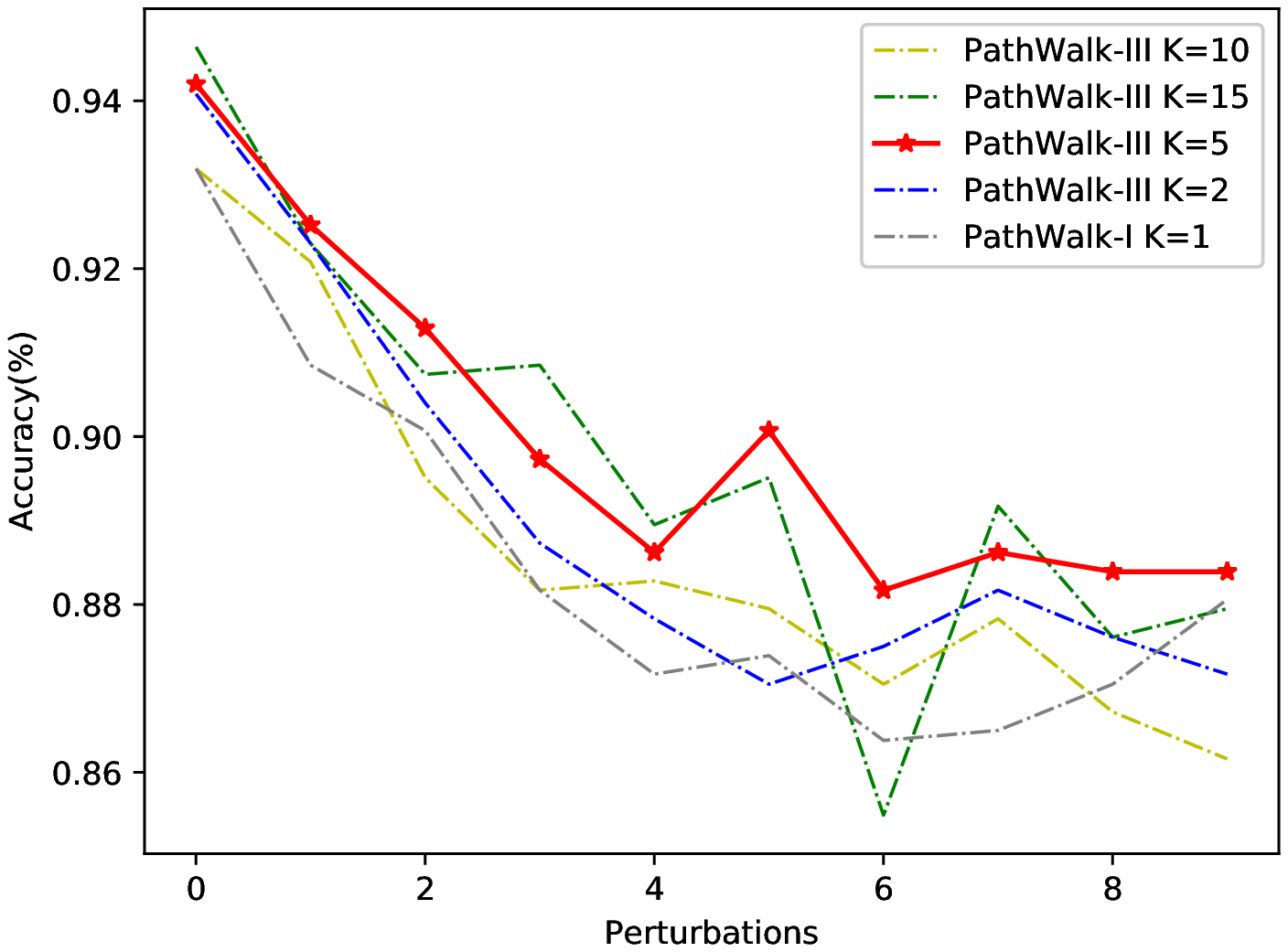}}
    \caption{Parameter sensitivity with varying $K$ from $1$ to $15$ on $DanMu$ dataset, Perturbations means testing accuracy when adding noise to the test dataset. a, PathWalk-$\uppercase\expandafter{\romannumeral2}$ method. b, PathWalk-$\uppercase\expandafter{\romannumeral3}$ with varying $K$} \label{fig_K_Sensitivity}
\end{figure}

The parameter $K$ is the number of topological features, it can affect the model's performance. In this subsection, when varying $K$ from $1$ to $15$, we perform experimental study of the influence of $K$ on the classification accuracy. Generally, a small number of topological features may lead to context sparsity, it enable us can not fully utilize information of external edges or external nodes from neighbors. Note that these features could bridge the gap on long range dependence inside $S$ as easily see from the Fig.~\ref{fig_overview}. While a large $K$ might lead to semantic drift.

Fig.~\ref{fig_K_Sensitivity} shows the broad trends of classification accuracy in decrease as increasing $K$.

Firstly, Fig.~\ref{fig_K_Sensitivity:a} shows the performance of PathWalk-$\uppercase\expandafter{\romannumeral2}$ is better at $K$ equal to $15$, it reflects that adding more external in-degree links to walk-of-words model as contextual information can improve accuracy. 

Secondly, however, the facts in Fig.\ref{fig_K_Sensitivity:b} are just the opposite, $K=5$ achieve better result than the others, note that the performance of $K=15$ is worse than $K=5$, it means adding more external in-degree links and external nodes as contexts that lead to semantic drift. $K=5$ is a tradeoff between contextual features and accuracy, so that we set $K$ equal to $5$ both for PathWalk-$\uppercase\expandafter{\romannumeral2}$ and PathWalk-$\uppercase\expandafter{\romannumeral3}$.

\subsection{Comparison with state-of-the-art}
In this subsection we compare PathWalk variants with the state-of-the-art models on four different domain datasets. All the methods are trained at the single word level. In particular, PathWalk variants are trained on term graph of single word, and the others also use bag of features at the single word level as input. 

Table~\ref{tab_accuracy} shows the comparison result. Firstly, PathWalk variants are usually better than the baselines in all cases, and achieve a strong benchmark results.

Secondly, on $DanMu$ dataset, PathWalk variants obtain higher performance gradually when introducing richer contextual information, but this consistent trend not appears in other datasets. In details, the $RT$ and $SST$ datasets contain much longer reviews, as listed in Table\ref{tab_datasets}, while the $DanMu$ dataset contains much shorter reviews among the four datasets. It is worth noting that PathWalk variants are much better on shorter sentences, and shows that many overlapped shorter texts in term graph are really complementary to each other. But, to longer sentences, although network topology is still helpful to improve the overall accuracy, the fine-tuning value of $K$ play a crucial role for classification. We leave how to select the best $K$ of topological features for expanding short sentence as future work.

In summary, our PathWalk method obtains the state-of-the-art performance in all four different domain datasets. The experiment result demonstrate that the non-local context could provide much gains to improve classification. Also it shows the size $K$ of external contextual features needs to be fine-tuned during training phase.

\begin{table}[htb]
\caption{Classification accuracy on four different datasets at the single word level, each test datasets contains $1000$ sentences.} \label{tab_accuracy}
\begin{tabular}{l|cccc}
\toprule
Method&$DanMu$&$RT$&$Query$&$SST$\\
\midrule
BiLSTM&94.08&61.61&93.08&74.22\\
fastText&92.70&54.10&92.70&51.85\\
TextCNN&92.40&58.70&91.10&70.09\\
\midrule
PathWalk-$\uppercase\expandafter{\romannumeral1}$&93.19&\textbf{62.50}&\textbf{94.08}&\textbf{75.39}\\
PathWalk-$\uppercase\expandafter{\romannumeral2}$&\textbf{94.20}&60.94&\textbf{93.53}&\textbf{77.21}\\
PathWalk-$\uppercase\expandafter{\romannumeral3}$&\textbf{94.53}&\textbf{61.72}&\textbf{94.31}&\textbf{75.39}\\
\bottomrule
\end{tabular}
\end{table}

\subsection{Comparison of Term Graph at Single Word vs. Word Segmentation}
Here we focus on the comparison between term graph at single word and word segmentation. For Chinese sentence dataset, we would like to know how performance is influenced by Chinese word segmentation. We apply a popular word segmentation, $jieba$ tool\footnote{https://github.com/fxsjy/jieba} to $DanMu$ and $Query$ datasets respectively, then we build graph using the result of $jieba$. All the methods are trained on word tokens level after word segmentation.

Table.~\ref{tab_tokensaccuracy} shows the comparison performance.

\begin{table}[htb]
\caption{Comparied Accuracy at the word token level after $jieba$ and at the single word level.} \label{tab_tokensaccuracy}
\begin{tabular}{l|c|c||c|c}
\toprule
\multirow{2}{*}{Method}&\multicolumn{2}{|c|}{Word Token Level}&\multicolumn{2}{||c}{Single Word Level}\\ \cline{2-5}
&$DanMu$&$Query$&$\emph{DanMu}$&$\emph{Query}$\\
\midrule
BiLSTM&96.32$\uparrow$&94.20$\uparrow$&\emph{94.08}&\emph{93.08}\\
fastText&91.00$\downarrow$&92.00$\downarrow$&\emph{92.70}&\emph{92.70}\\
TextCNN&96.10$\uparrow$&93.00$\uparrow$&\emph{92.40}&\emph{91.10}\\
\midrule
PathWalk-$\uppercase\expandafter{\romannumeral1}$&95.09$\uparrow$&\textbf{94.64}$\uparrow$&\emph{93.19}&\emph{94.08}\\
PathWalk-$\uppercase\expandafter{\romannumeral2}$&\textbf{96.54}$\uparrow$&\textbf{95.31}$\uparrow$&\emph{94.20}&\emph{93.53}\\
PathWalk-$\uppercase\expandafter{\romannumeral3}$&\textbf{96.65}$\uparrow$&94.64$\uparrow$&\emph{94.75}&\emph{94.31}\\
\bottomrule
\end{tabular}
\end{table}

Firstly, as we can see, the overall accuracy is increased but the $fastText$ method in decrease slightly, as seen the symbol $\uparrow$ and $\downarrow$ tagging. It shows word segmentation are helpful compared with single word. The reason is that the word token is meaningful, but the drawback is their feature space become more sparse.

Secondly, our PathWalk is often better than the three baselines. Note that PathWalk variants gradually improve the accuracy in a consistent way across the two datasets, from $95.09\%$ up to $96.65\%$ and from $94.64\%$ to $95.31\%$ respectively, but PathWalk-$\uppercase\expandafter{\romannumeral3}$ on $Query$ dataset drops to $94.64\%$ since introducing more contexts lead to semantic drift.

\subsection{Adversarial experiment: perturbation on test phase}
The quality of robustness can be measured in adversarial experiment, in this work we adopt two types of adversarial experiment, the first is only adding noises to test dataset, while the second is attacking the model during training phase via adding noises to training dataset. In this subsection, we perform the first experiment at the single word level. 

How small noises attack on the content of test sentence could affect the classification performance. We randomly sample words as noise padding to the test sentence, where the noises come from vocabulary of respective training datasets. However in this case, we assume that the perturbed sentence will preserve semantics of the original, because of a sentence size is very smaller than the size of vocabulary, so that it is less likely to change the meaning of the original sentence.

To test robustness of the pre-trained model, we gradually increase the number of noise words, with varying from $1$ to $8$, this is equivalent to directly attack the pre-trained model by unseen sentences, and then observe their robustness to noises in terms of classification accuracy. 
\begin{figure}[htb] \centering
    \subfigure[]{
    \label{fig_danmu:a}
    \includegraphics[width=0.22\textwidth]{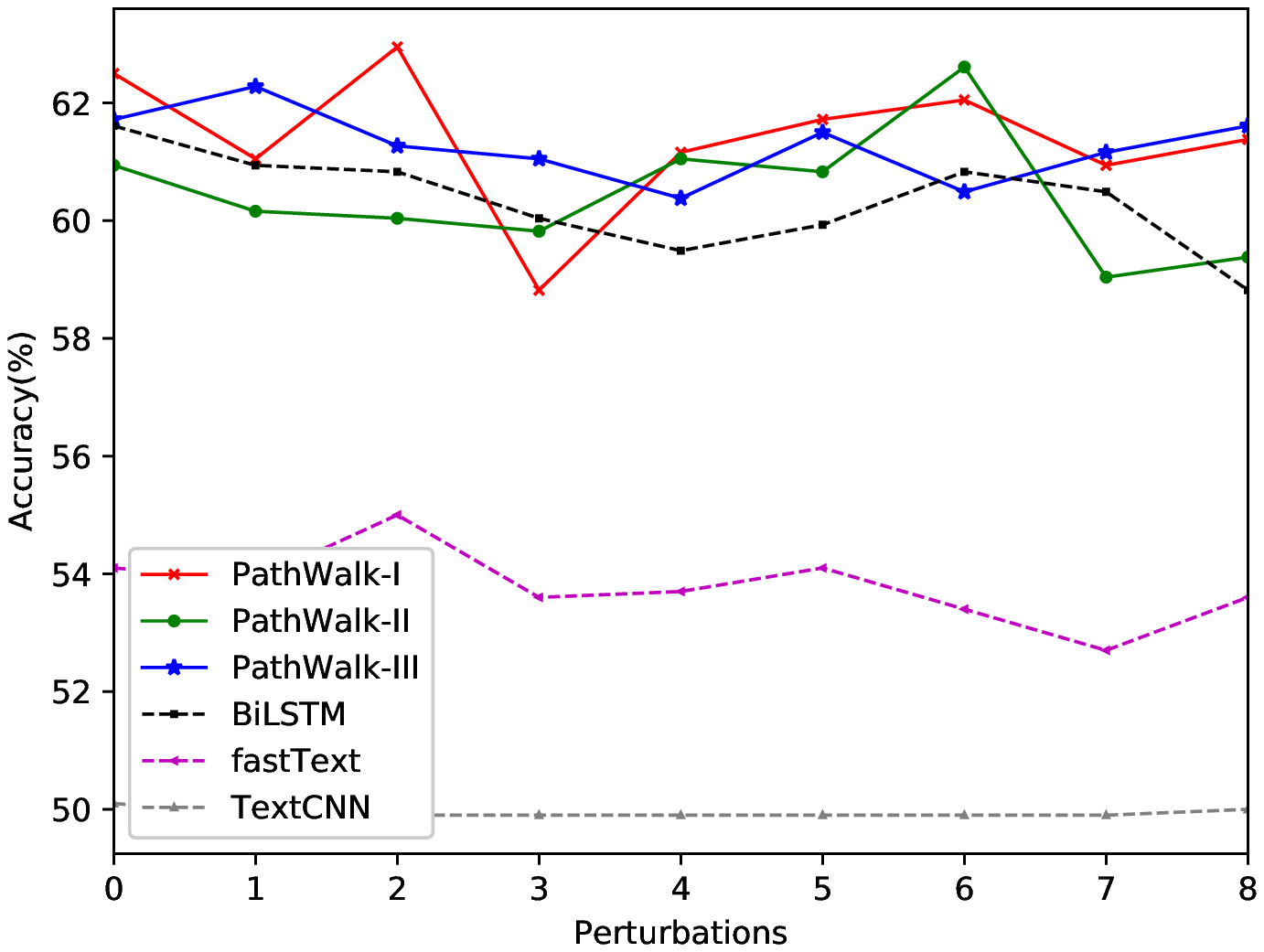}}
   \vspace{0.00in}
   \subfigure[]{
    \label{fig_danmu:b}
    \includegraphics[width=0.22\textwidth]{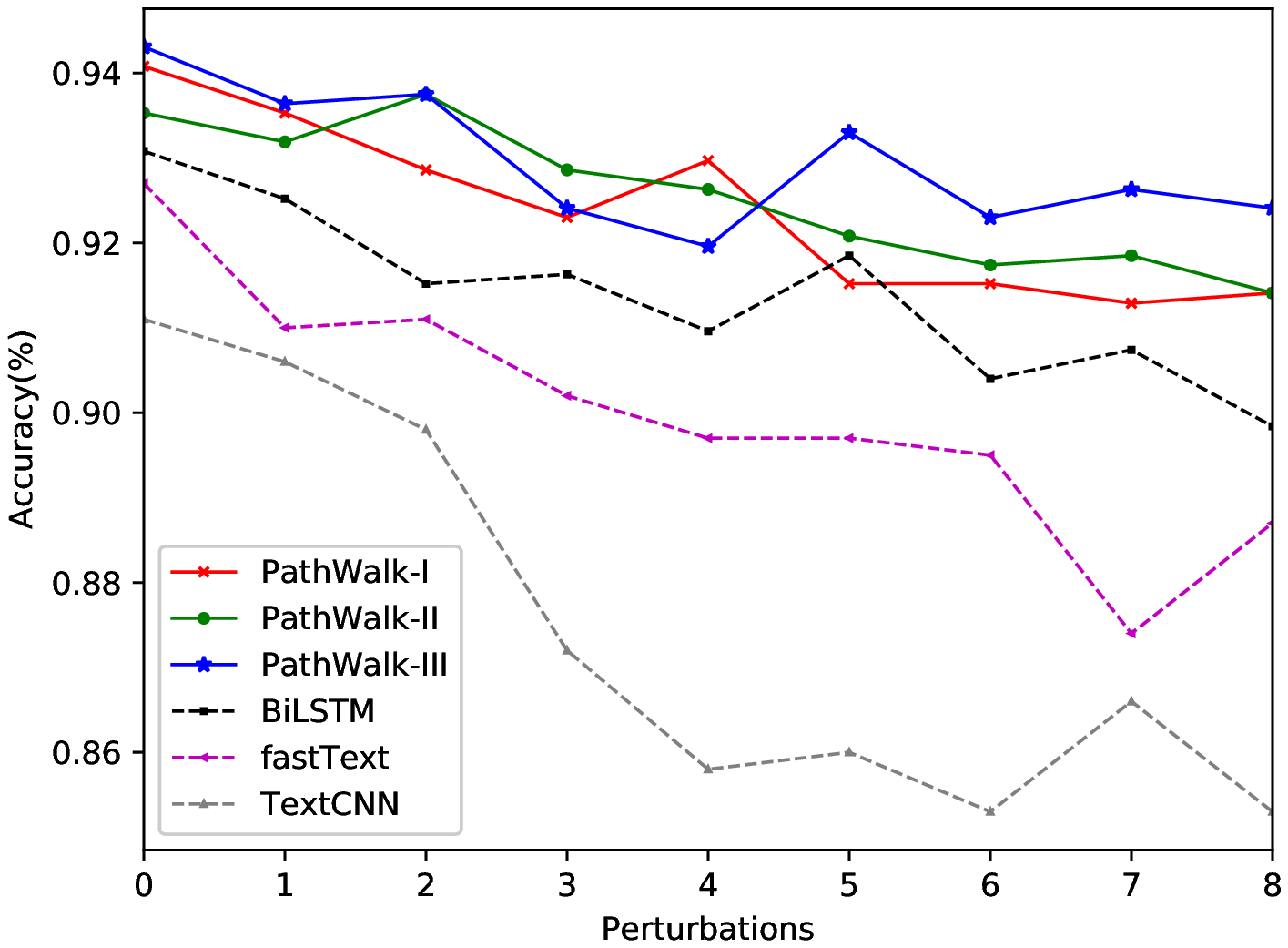}}
    \caption{Adversarial perturbation on $RT$ and $Query$ dataset with increasing number of perturbations. a, experimental result on English $RT$ dataset. b, on Chinese Video Query dataset. The noise word we randomly select from each Vocabulary.} \label{fig_danmu}
\end{figure}

Firstly, Table~\ref{tab_testperturbation} shows perturbation result on the two different domain dataset, i.e. $DanMu$ and $SST$. When gradually increasing noise words, the overall accuracy result in decrease. Surprisingly our PathWalk variants shows more stable than others, and importantly result in a lower decrease in accuracy when increasing noises compared with the baselines.

Secondly, Fig.~\ref{fig_danmu} shows the results on $RT$ and $Query$ datasets. Fig.\ref{fig_danmu:b} shows the accuracy of PathWalk variants are more stable than the others on $Query$ dataset, and also consistently exceeds the baselines. In contrast, Fig.\ref{fig_danmu:a} shows the accuracy of PathWalk-$\uppercase\expandafter{\romannumeral2}$ is worse than the other two PathWalk variant and $BiLSTM$ method. However, all the methods on $RT$ dataset achieve a lower accuracy, due to the standard deviation of $RT$ sentence length is greater than the other three datasets, in this case it indicate that our strategy of fixed-sentence length is not appropriate for dataset with high variance.

\begin{table*}[htb]
\centering
\caption{Adversarial perturbation on $DanMu$ and $SST$ dataset with increasing number of perturbations, $+i$ means we add $i$ noise words to every sentence of both dataset.} \label{tab_testperturbation}
\begin{tabular}{l|ccc|ccc}
\toprule
\multirow{2}{*}{Test}&\multicolumn{3}{|c}{PathWalk}&\multicolumn{3}{|c}{Baselines}\\ \cline{2-7}
&$\uppercase\expandafter{\romannumeral1}$ & $\uppercase\expandafter{\romannumeral2}$&$\uppercase\expandafter{\romannumeral3}$&BiLSTM&fastText&TextCNN\\
\midrule
$DanMu$&93.19&94.20&94.53&94.08&92.90&92.40\\
$+1$&90.85&91.63&91.74&90.62&92.40&87.10\\
$+2$&90.07&90.51&88.84&89.51&90.50&81.20\\
$+3$&88.17&89.17&88.06&87.83&89.70&75.00\\
$+4$&87.17&87.39&87.61&86.94&89.00&71.10\\
$+5$&87.39&86.83&86.72&87.17&89.10&67.70\\
$+6$&86.38&86.72&86.83&87.95&88.30&64.30\\
$+7$&86.50&87.39&86.38&84.60&85.90&63.10\\
$+8$&87.05&86.50&87.83&83.93&85.30&63.40\\
\bottomrule
$SST$&75.39&77.21&77.08&74.22&51.85&70.09\\
$+1$&74.74&76.30&75.26&73.96&51.73&68.36\\
$+2$&75.39&76.04&75.26&73.31&52.66&66.63\\
$+3$&73.31&75.00&75.91&73.57&51.62&65.47\\
$+4$&74.87&76.17&74.35&73.57&-&65.47\\
$+5$&73.44&74.09&75.52&73.05&-&65.13\\
$+6$&74.09&75.26&73.57&73.05&-&63.74\\
$+7$&73.57&76.17&75.52&73.07&-&64.43\\
$+8$&73.83&74.74&74.22&72.53&-&62.82\\
\bottomrule
\end{tabular}
\end{table*}

Overall, our methods outperform the strong baselines in robustness on three out of four datasets, except $RT$ since it's fixed-sentence rule is not work well.

\subsection{Adversarial experiment: perturbation during model training}
Prior experiments we focus on robustness measure during test time. In this subsection we focus on the second adversarial experiment, attacking the model during training phase. Here we only conduct experiment on walk-of-words model using PathWalk-$\uppercase\expandafter{\romannumeral1}$ method, through adding randomly perturbations to each training sentence before fed it into model. 

In details, the adversarial noises are directly added to each word of target sentence, where the noise comes from the term graph itself, including node perturbation or edge perturbations, then the sentence representation can be written as:
\begin{equation} \label{equ_attack}
\overrightarrow{S}_{1:n} = \sum_{i=1}^{n}\{\overrightarrow{v}_{i} + \overrightarrow{ie}_{i} + \varepsilon \times \overrightarrow{o}^{(i)}\},
\end{equation}
where $\varepsilon$ is a hyperparameter and $\overrightarrow{o}$ is an embedding of the sampled edge or node in term graph as noises during training. Also, the short sentence model can also be interpreted as walk-of-words model plus the adversarial perturbations for each word.

\begin{table}[htb]
\caption{Adversarial training on $DanMu$ dataset using graph of single word, the different is the sentence representation is changed to Eq.(\ref{equ_attack}), where $\varepsilon = 1.0$. Detailedly, the $^{\ell}$ labeled method only add one edge as noises to each word of target sentence, similarly, the $^{\ast}$ method add four edges as noises, the $\wr$ add four edges and nodes as noises. Where all the small perturbations are derived from term graph itself. $+i$ is the same as before.} \label{tab_Adversarial}
\begin{tabular}{l|ccc|c}
\toprule
\multirow{2}{*}{Test}&\multicolumn{4}{|c}{PathWalk}\\ \cline{2-5}
&$\uppercase\expandafter{\romannumeral1}^{\ell}$ & $\uppercase\expandafter{\romannumeral1}^{\ast}$&$\uppercase\expandafter{\romannumeral1}^{\wr}$&$\uppercase\expandafter{\romannumeral1}$\\
\midrule
$DanMu$&95.42$\uparrow$&94.64$\uparrow$&93.30&\emph{93.19}$^{\dagger}$\\
$+1$&94.08&93.08&87.83$\downarrow$&\emph{90.85}\\
$+2$&91.96&91.29&86.50&\emph{90.07}\\
$+3$&89.51&91.85&83.04&\emph{88.17}\\
$+4$&91.41&90.18&83.82&\emph{87.17}\\
$+5$&89.62&89.62&79.69&\emph{87.39}\\
$+6$&89.73&90.51&79.58&\emph{86.38}\\
$+7$&88.28&89.06&78.46&\emph{86.50}\\
$+8$&89.29&88.28&75.11&\emph{87.05}\\
\bottomrule
\multicolumn{4}{l}{$^{\dagger}$ the result of Table.\ref{tab_testperturbation} as baseline.}
\end{tabular}
\end{table}

Table~\ref{tab_Adversarial} shows the robustness performance, the last \emph{column} as our baseline. 

As seen, PathWalk-$\uppercase\expandafter{\romannumeral1}^{\ell}$ that attacked by only one edge perturbation, and achieves the significant improvement. Also surprisedly, PathWalk-$\uppercase\expandafter{\romannumeral1}^{\ell}$ outperforms PathWalk-$\uppercase\expandafter{\romannumeral1}$, including both of the single word level and word segmentation level. In $DanMu$ sentiment classification task, the PathWalk-$\uppercase\expandafter{\romannumeral1}^{\ell}$ by adversarial training improves the overall accuracy of the original PathWalk-$\uppercase\expandafter{\romannumeral1}$ at single word level from $93.19\%$ to $95.42\%$. PathWalk-$\uppercase\expandafter{\romannumeral1}^{\ast}$ also gives better results than the baseline. While, PathWalk-$\uppercase\expandafter{\romannumeral1}^{\wr}$ perform worse than the baseline, because of it suffer many times perturbations. 

In summary, adversarial training is an interesting problem and the result shows that a small perturbations are efficient in improving the model's performance.

\section{Conclusions}
In this work, we replace IID dataset assumption with Non-IID to perform short text classification more efficiently. Under this Non-IID, we use graph networks of words to represent the whole training corpus. In the graph, a short sentence can capture more contextual information than in its own self, and in other words, it alleviate the problem of data sparsity for short text. Experiment on four different domain datasets show that term graph and network topology could improve performance in classification accuracy and robustness. When a short sentence attacked with small noises, our PathWalk method is proved to be more stable than the baselines.

In the future, we are going to explore the use of topological features based on attention methods.

\section{Acknowledgements}
The idea of the paper had its beginning in Dec. 2017, and the paper was written in May 2018. I don't have any time to revise it, so upload it to arXiv. I'm currently a PhD at the University of BUPT. My research is centered around the problem of vision and language, including visual dialogue, vision grounding, visual question generation, visual reasoning, referring expression and natural language processing. If you any questions, please contact me with pangweitf@bupt.edu.cn or pangweitf@163.com.

\bibliographystyle{aaai} \bibliography{sample-bibliography.bib}

\end{document}